# Demonstration of Stimulated Supercontinuum Generation – An Optical Tipping Point

D. R. Solli[1], C. Ropers[1,2] & B. Jalali[1]

[1]*Department of Electrical Engineering, University of California, Los Angeles, 90095 USA;* [2]*Max Born Institute for Nonlinear Optics and Short Pulse Spectroscopy, D-12489 Berlin, Germany*

**Abstract:**

Optical supercontinuum radiation, a special kind of white light, has found numerous applications in scientific research and technology. This bright, broadband radiation can be generated from nearly monochromatic light through the cooperative action of multiple nonlinear effects. Unfortunately, supercontinuum radiation is plagued by large spectral and temporal fluctuations owing to the spontaneous initiation of the generation process. While these fluctuations give rise to fascinating behavior in the form of optical rogue waves[1], they impede many critical applications of supercontinuum. Here, we introduce, and experimentally demonstrate, a powerful means of control over supercontinuum generation by stimulating the process with a very weak optical seed signal[2]. This minute addition significantly reduces the input power threshold for the process and dramatically increases the stability of the resulting radiation. This effect represents an optical tipping point, as the controlled addition of a specialized, but extraordinarily weak perturbation powerfully impacts a much stronger optical field, inducing a drastic transition in the optical system.

Nonlinear processes have occupied a central role in physics during the last half-century. Especially in optics, the availability of high-intensity, short pulses of coherent radiation have brought nonlinear effects to center stage. In addition, the rise of computers, especially desktop machines, has facilitated the modelling and understanding of many counterintuitive effects that appear in nonlinear systems. A particularly striking example of nonlinear action in optics is the generation of extremely broadband "supercontinuum" radiation using comparatively narrowband light as the only ingredient. First discovered in the late 1960s by Shapiro and Alfano[3], the spectral width of supercontinuum (SC) light can span multiple optical octaves in its extreme forms[4]. This broadband radiation differs markedly from the white light produced by an ordinary light bulb because it can be far brighter, more coherent, and directional. SC white light has many applications including optical frequency metrology[5], spectroscopy[6], optical coherence tomography[7], and data transmission[8], to name just a few.

A multitude of nonlinear optical effects contribute to the generation of SC. The relative contributions of these effects depend dramatically on numerous factors, including the characteristics of the starting radiation, as well as both the linear and nonlinear properties of the optical medium. Moreover, a unified explanation has been formulated only relatively recently[4], and owing to the counterintuitive hallmark of nonlinear effects, additional surprises no doubt remain. The primary method for producing ultra-broadband SC involves the so-called soliton-fission mechanism, a process that occurs when intense pump pulses are launched into a nonlinear optical fiber at its zero-dispersion wavelength. In this situation, the input light fissions into multiple solitons—nonlinear wavepackets that propagate without spreading due to a balance between dispersion and nonlinearity[9]—at different red-shifted wavelengths, along with blue-shifted dispersive radiation[10].

Unfortunately, noise plays a significant role in SC generation: the combination of dispersion and nonlinearity amplifies the noise present in the input light, which can seriously degrade the stability of the SC[11,12,13,14]. This instability comprises significant spectral fluctuations, loss of coherence, and other forms of jitter[4]. For example, single-shot measurements have revealed variable fine-scale structure in individual SC pulses that escapes unnoticed when many pulses are averaged together[15]. These instabilities negatively impact many important applications of SC, such as optical frequency metrology[16], generation of ultrashort optical pulses[17], photonic time-stretch analog-to-digital conversion[18], and optical coherence tomography[19]. Optical coherence tomography, for example, an important biomedical imaging technique, requires a low-cost source of broadband radiation, but loses sensitivity because of intensity noise[20].

At the heart of the initiation of soliton-fission based SC generation is a nonlinear growth process known as modulation instability (MI)[9], an effect that adds sidebands to narrow-band input radiation. These sidebands increase the optical bandwidth of the input pulse to the point where the Raman-induced self-frequency shift[9] can separate it into red-shifted solitons[21,22,23]. Because the MI growth starts spontaneously from background noise and amplifies its influence[11,24], the subsequent initiation of soliton fission becomes inherently unpredictable[4]. Particularly for longer input pulses, much of the initial spectral broadening preceding soliton fission is spontaneously initiated by either shot noise, spontaneous emission, or other source fluctuations[11,12,13,14]; as a result, the resulting SC can become nearly incoherent away from the pump wavelength, and is subject to large pulse-to-pulse spectral variations[4,22,25]. Just below the soliton-fission power threshold, this variability can be so extreme that rare, but bright flashes of white light—optical rogue waves—are produced[1]. These studies also showed that rogue waves are produced by accelerated SC generation—induced by a particular type of noise in the input—showing a path to stimulating their generation.

Here, we demonstrate that SC generation can be controlled by applying a tailored, but extremely weak coherent seed pulse. The weak seed pulse stimulates SC generation by initiating modulation instability with a controlled signal rather than noise. We show that this seed pulse lowers the input (pump) power threshold and reduces the noise-induced instability of the resulting white light. Induced MI has been used to generate high-repetition rate trains of soliton pulses[26,27], promote the formation of Raman solitons[28,29], and produce a MI-soliton laser[30]; however, the enhancement and stabilization of SC generation with a weak stimulus has not yet been demonstrated.

SC seeding resembles the manner in which a nucleation point facilitates the initiation of boiling in hot water. Without nucleation sites, water must be superheated to initiate boiling, which results in a spontaneous, uncontrolled initiation process. Similarly, conventional (unseeded) SC generation has a larger power threshold and begins randomly[4]. Like a pot of nearly boiling water, light at the cusp of SC generation is at a tipping point: a minute perturbation can have a powerful effect on the system, stimulating the onset of a dramatic transition. In the time domain, this tipping point also has a striking manifestation: whereas unseeded sub-threshold pump pulses generate only rare flashes of SC radiation, introduction of the seed causes the once rare events to become the predictable norm. In our experiments, we employ an accurately timed, frequency-shifted, weak seed pulse (only 0.01% of the pump intensity), which experimentally lowers the SC power threshold by more than 25%, produces optical switching with a dynamic range of 30 dB, and reduces the pulse-to-pulse standard deviation of power spectral density by 90%. The seed also increases the pulse-to-pulse coherence of the output light dramatically.

The master optical source for our experiments is a mode-locked laser producing near-transform limited picosecond pulses with a wavelength of 1550 nm and a repetition rate of 25 MHz. The primary portion of the laser output is spectrally filtered to a bandwidth of approximately 1 nm and amplified using a large-mode-area erbium-doped fiber amplifier capable of producing pump pulses with relatively high peak power. The amplified pulses are approximately 3.7 ps in duration, as determined by autocorrelation. These amplified pulses are directed into a highly nonlinear optical fiber (15 m length, Corning HNL ZD 1550) with very low dispersion at 1550 nm in order to produce SC. A minute portion of the mode-locked laser output is directed into an independent optical amplifier, which spectrally broadens the input light for use as the seed pulse. This broadened light is frequency filtered to select a weak portion centered near 1630 nm (the spectral region where the nonlinear fiber produces MI). Using an optical delay stage, the coherent seed (0.01% of the pump intensity) is precisely timed and delivered to the highly nonlinear fiber along with the pump. As illustrated schematically in Figure 1A, the output light depends dramatically on the timing of this weak seed.

At the output of the nonlinear fiber, we measure the spectrum of the output radiation with a conventional spectrometer. In Figure 1B, we show the output spectra for pump power levels below and above threshold without the proper timing of the weak seed. Below threshold, the central portion of the spectrum shows the characteristic undulations of self-phase modulation of the input light (fine oscillations observed on the central peak), along with broad MI lobes[9]. As the input power is increased, the central portion of the spectrum continues to broaden, and the power within the MI lobes increases until a soliton is ejected from the red MI lobe. Once a soliton pulse splits off from the long-wavelength MI lobe, it shifts towards red wavelengths due to stimulated Raman scattering[4]. The soliton ejection marks the beginning of soliton fission, and leads to significant spectral broadening. When the weak seed is not properly timed, the results are essentially identical to what is obtained without any seed at all. On the other hand, the situation changes dramatically when the weak seed is properly timed with the pump (cf. Figure 1C). At low pump power, the MI lobes become much sharper, and the power threshold for soliton fission is now much lower.

These spectra clearly show that SC generation is dramatically accelerated by the addition of an extremely weak seed pulse. To quantify the spectral width of the SC, we carve out a portion of the SC spectrum $(\lambda \approx 1685 \text{ nm}, \Delta\lambda \approx 10 \text{ nm})$ beyond the red edge of the seed pulse. We monitor the output power within this wavelength band as a function of the pump power, with and without the properly timed seed. Figure 2A shows that this red-shifted output power exhibits a sharp transition threshold, and that this threshold is significantly reduced by the presence of the seed. Moreover, the addition of the seed at a fixed pump power creates a power contrast of 30 dB, an effect that can be utilized for high-contrast optical switching. In essence, an extremely weak seed has a dramatic influence on a pump pulse that is 10,000 times more intense.

When the seed stimulates SC emission, the pulse-to-pulse variability is reduced. Because pulse-to-pulse variations are so rapid, however, a conventional spectrometer cannot measure them; even a wildly fluctuating SC typically appears relatively well behaved on a spectrometer, which measures only its time-averaged spectrum. Here, we measure the spectral variations of the SC in real time using group-velocity dispersion (GVD) to temporally stretch the pulses so that a large number of ultra-short events can be captured with a single photodetector and a single-shot oscilloscope. This wavelength-to-time transformation technique, which was also recently used to identify optical rogue waves[1], permits real-time spectral measurements and allows us to measure the variation in the spectral width of the SC from pulse to pulse.

Using the wavelength-to-time transformation technique, we demonstrate that a weak perturbation can be used to reduce unwanted fluctuations in SC radiation. In

Figure 2B, the statistical histograms of bandpass-filtered SC pulses with and without the properly-timed weak seed are contrasted. The pump power was adjusted to keep the average SC output power at the same level in both cases. This measurement shows a dramatic 90% reduction in the standard deviation of pulse-to-pulse intensity variations. Furthermore, the fluctuations clearly depend on the relative delay between the pump and seed (cf. Fig. 2C): when the seed is not present only rare, unpredictable SC pulses are produced. These measurements prove that the seeding process facilitates controlled SC generation from a pump pulse that would otherwise have insufficient intensity to produce significant SC. Moreover, we expect that an even greater stability than shown here can be reached with this approach, if the current limitations from residual seed fluctuations are reduced.

To analyze the problem in more detail, we also numerically model SC generation in the presence of a weak seed pulse. The generalized nonlinear Schrödinger equation (NLSE) includes the effects of dispersion, the Kerr nonlinearity, and Raman scattering. Using the NLSE, we calculate the output field as a function of the relative time delay (cf. Figure 3) and frequency shift between the pump and seed pulses, and quantify the amount of broadening by integrating the power spectral density over red-shifted wavelengths $(\lambda > 1685\,\text{nm})$. We use a pump pulse that has insufficient power to generate SC without the seed, and to include the effects of input noise, we add a small amount of bandwidth-limited random noise to its envelope. Although the precise form of this stochastic perturbation is not critical, here we choose a simple form with amplitude proportional to the instantaneous field strength. As illustrated in the maps of Figure 3, this calculation displays a well-defined "sensitive spot" of time-frequency shifts over which the seed induces significant SC output. Outside of this vicinity, the impact of the seed is negligible by comparison.

One may wonder why a particular temporal region on the envelope of the input pulse is most sensitive to perturbation. A previous study has observed that spontaneous MI can lead to the growth of disturbances on the steep regions of a pulse envelope[21]; however, under the conditions studied, spontaneous growth appeared on both sides of the envelope, and the initial formation was attributed to frequency beating. In the present situation, some insight can be found by studying the pulse dynamics in the numerical model. In the initial stages of propagation, Raman scattering is negligible, and modifications to the pump field are produced primarily by the Kerr nonlinearity, which is responsible for both self-phase modulation (SPM) and MI. SPM imparts a frequency chirp to the pulse by shifting the carrier frequency on its leading and trailing slopes, where the intensity slew rate is largest[9]. This broadens the pulse spectrum, distributing downshifted frequency components on its leading edge and upshifted components on its trailing edge. Consequently, the leading edge shifts away from the zero-dispersion wavelength and into the fiber's anomalous dispersion regime; this portion now drives MI, whereas, the trailing edge does not because it experiences normal dispersion. Furthermore, the MI frequency shift for the components on the leading edge decreases with greater anomalous dispersion. This enhances the effective MI interaction because the group velocity of the MI disturbance matches that of its corresponding pump component more closely. In other words, if the MI shift is too large, the disturbance cannot grow effectively because it "walks off" rapidly from the appropriate region of the pump pulse.

The trade-off creates a special region on the pump envelope where the MI gain is large enough and the walk-off is not too rapid. A properly timed seed with frequency matching the MI shift of this special region will experience the largest cumulative MI gain. It is important to note that because the SPM and MI are occurring simultaneously during the propagation, the two effects cannot be analyzed independently, and the precise frequency shift and timing of the effective seed are not intuitively obvious. As seen in Figure 3C, other effects such as spectral interference can also lead to additional structure within the special region. Nevertheless, numerical simulations show the effect of the seed in striking fashion. Once the cumulative gain becomes large enough, soliton fission begins suddenly as a large disturbance, travelling much slower than the remnants of the original envelope. This disturbance rapidly disintegrates the original pulse shape, and liberates one or more narrow solitons while leading to rapid spectral broadening. Without the seed, soliton fission will occur if the pump power is large enough; however, the process begins rather suddenly and explosively (like the onset of boiling in a superheated liquid), typically fissioning into multiple solitonic pulses. On the other hand, when the seed is present, a less intense pump (that generates SC with equivalent average power) produces a soliton at an earlier stage of propagation.

We also model the impact of the weak seed on the SC amplitude fluctuations and phase coherence. As mentioned above, we include noise on the input pulse envelope to reproduce the random properties normally attributed to SC radiation. We then solve the NLSE repeatedly for many independent events with and without the weak seed, and spectrally filter the output to collect the light within a red-shifted frequency band. To illustrate the impact of the seed on both the amplitude stability and phase coherence, we present a complex-plane scatter plot of the electric field from independent events (cf. Figure 4A). Above threshold, the unseeded scatter fills a large region of the complex plane, whereas, the seeded scatter remains within a tiny region. Clearly, the seed produces a strong SC with much more stable amplitude and phase. As the pump power is increased, it becomes more sensitive to seed fluctuations (or noise); thus, from an experimental perspective, a more stable seed pulse is needed to stabilize SC generation with a higher power pump. For comparison with the stability plot, Figure 4B shows representative single-shot SC spectra with and without the seed at a fixed (low) pump power. By varying the pump power, we also calculate the SC threshold (quantified by the red-shifted power as previously described) with and without the seed, as shown in Figure 4C.

As a side note, we point out that stabilized, ultrashort pump pulses (< 50 fs) are used to produce coherent SC for applications such as optical frequency metrology, and studies have shown that SC coherence increases significantly as the pulse width is reduced[31]. For such pulses, the bandwidth is so great that MI may be seeded by the tail of the spectrum, which is a likely factor in the enhanced SC stability[4]. However, here we highlight the significance of deliberate seeding, which offers much greater control over the effect, and also extends the benefits to SC generation with longer pulses. Indeed, ultrashort pulses and the sources that generate them are not practical in many situations, so enhancing the stability of SC from longer pulses is an important problem. Control over the seeding process and optimization of the seed pulse may also add further improvement to the stability of SC generation with ultrashort pulses. Specialized dispersion-decreasing fibers[11] and dispersion-managed fibers[23], as well as soliton pulse compression[11,32] can also be used to reduce excess noise, particularly with longer pulses; however, the present method substantially expands the repertoire of possibilities, and may also be used in conjunction with other approaches.

In conclusion, SC generation is characterized by an extremely sharp power transition threshold. Just below threshold, very little white-light average power is generated, although short, rare bursts of intense white light may be spontaneously emitted due to accidental seeding of MI by random noise. On other hand, if a weak, tailored perturbation co-propagates with the pump pulses, intense SC pulses become the norm. In other words, this system exhibits a tipping point: the addition of a weak perturbation increases the SC output dramatically by causing an event that is initially exceedingly rare to become the most probable result. The presence of a tipping point allows one to exert significant control over a system with minute effort, and in the present case, offers a means to increase the efficiency of SC generation and stabilize its properties. Because an exceedingly weak field is used to control a much stronger one, this phenomenon may also lead to new applications of SC, such as high-contrast all-optical switching. Finally, we point out that this effect may also be useful in influencing the dynamics of other nonlinear systems.

Supplementary Material

Supercontinuum (SC) generation can be modelled with the generalized nonlinear Schrödinger equation (NLSE), which describes the evolution of light waves propagating in optical fiber—a dispersive, nonlinear medium[9]. Specifically, the NLSE governs the dynamics of the slowly-varying envelope of a pulse or wavepacket under the influence of dispersion, as well as the Kerr nonlinearity, self-steepening, and the vibrational Raman response of the fiber. Formulated in terms of the reference frame co-moving with the pulse, the NLSE can be written in a relatively simple form:

$$\frac{\partial A}{\partial z} - i\sum_{m=2} \frac{i^m \beta_m}{m!} \frac{\partial^m A}{\partial t^m} = i\gamma \left[ |A|^2 A + \frac{i}{\omega_0} \frac{\partial}{\partial t}\left(|A|^2 A\right) - T_R A \frac{\partial |A|^2}{\partial t} \right],$$

where $A = A(z,t)$ is the slowly-varying electric field envelope, $\beta_m$ are values that characterize the fiber dispersion, $\gamma$ is the nonlinear coefficient of the fiber, $\omega_0$ is the central carrier frequency of the field, and $T_R$ is a parameter that characterizes the Raman response of silica fiber. The second and third terms on the right-hand side are approximations that describe self-steepening and the vibrational Raman response of the medium; these approximations have been successfully used to model SC generation in the presence of input noise, and they lend themselves to efficient numerical solution with the split-step method.

The Kerr nonlinearity produces self-phase modulation (SPM), and the Raman term causes frequency downshifting of the carrier wave. The combined action of the Kerr and the dispersive terms produce modulation instability (MI), a four-wave mixing process in which energy is transferred to sidebands on both sides of the pump spectrum. Normally, MI arises for signals in the anomalous dispersion regime $(\beta_2 < 0)$, and does not occur in the normal dispersion regime $(\beta_2 > 0)$. If a narrowband pulse is launched into a fiber near its zero-dispersion wavelength $(\beta_2 \approx 0)$, MI may still be generated due to higher order dispersion, although the MI frequency shift increases as the total dispersion is reduced. When the MI frequency shift is large, the transfer of energy to the sidebands loses efficiency because the new frequency components quickly separate or "walk-off" from the pump due to mismatched group-velocities.

On the other hand, the role of MI with pulses is complicated by spectral broadening. In particular, the spectrum of the input light will broaden due to self-phase modulation, an effect that arises because the time-dependent intensity profile of the pulse alters its own optical phase through the Kerr nonlinearity. This effect adds new frequency components to the pulse envelope wherever it has a nonzero temporal derivative:

$$\Delta\omega = \frac{d}{dt}\Delta\phi \propto \frac{d}{dt}|A|^2.$$

In words, the SPM imparts a frequency chirp to the pulse: the carrier frequency is downshifted on the leading slope of the envelope, upshifted on the trailing slope, and unchanged in the center where the time-derivative is zero. The effect of SPM is especially strong for a pulse launched at the zero-dispersion wavelength because the pulse envelope does not disperse quickly. The spectral broadening shifts the leading edge of the pulse into the anomalous dispersion regime, and these components begin to excite MI with a reduced frequency shift, and as a result, these new components are able to excite MI more efficiently. However, as the optical intensity on the pulse's leading edge is obviously lower than at the peak, the instantaneous MI gain for this portion of the pulse is smaller. This leads to a trade-off between gain and walk-off, which favors the growth of a disturbance on a limited range of the envelope at the corresponding MI shift; this disturbance has the greatest time-integrated MI gain. Furthermore, a seed launched with this frequency and timing will grow more rapidly than one with other parameters. A quantitative prediction of the frequency shift and timing of the ideal seed requires simultaneous consideration of the broadening due to SPM, walk-off, and the MI frequency shift and gain, which are all interconnected.

In our calculations, we include the second and third order dispersion of the fiber used in our experiments, as provided in the manufacturer's test data (Corning HNL ZD 1550). At an operating wavelength of 1550 nm, we have: $\beta_2 \approx 1.13 \times 10^{-4}$ ps$^2$/m and $\beta_3 = 6.48 \times 10^{-5}$ ps$^3$/m. Although inclusion of additional higher-order dispersion terms may facilitate precise calculations, they are not required to corroborate our experimental observations. The nonlinear coefficient and the Raman response parameter are given by: $\gamma = 10.66$ W$^{-1}$km$^{-1}$ and $T_R = 5$ fs. The pulse duration is chosen to be approximately 3.7 ps (transform limited) in accordance with the experiment.

We include random noise in our calculations by adding a small randomly chosen number to the input field envelope at each point in time. The amplitude of the random distribution at each point is proportional to the instantaneous amplitude of the pulse. We then apply a frequency bandpass filter to limit the input noise to a bandwidth of approximately 30 THz around the seed wavelength. The overall noise amplitude influences the power threshold for SC generation by soliton fission; we choose the overall noise amplitude so that the threshold is similar to what we observe in the experiment without the weak coherent seed. Since unseeded soliton fission is randomly initiated by noise and can vary significantly from pulse to pulse, we typically calculate many events to determine the average SC at each power level. We have observed that the qualitative effects described in the manuscript do not depend on a specific choice of the noise perturbation, but we choose the present form merely because it is convenient for efficient numerical computation and is physically intuitive.

To model stimulated SC generation, we add a weak, coherent seed to co-propagate with the input pulse in the nonlinear fiber. To calculate the ideal seed frequency shift and timing, we use a 200 fs (transform-limited) seed pulse, and vary its center frequency and time delay relative to the pump pulse. In each case, we integrate the output spectrum over red-shifted wavelengths (>1680 nm), and we then compare the magnitudes to determine the "sensitive spot" for the seed. We find that a seed strength of less than 0.0001% of the pump intensity is sufficient to induce appreciable SC generation; however, the weaker the seed, the smaller the sweet spot becomes.

In the experiment, the seed pulse intensity is roughly 0.01% of the pump, and its temporal duration is comparable to that of the pump; when modelling the stimulated SC threshold and stabilization effect, we use these parameters. In order to quantify the spectral amplitude and phase fluctuations, we calculate the propagation for many independent events with and without the seed, integrate the output field over a 5 nm interval centered at roughly 1950 nm (~400 nm red-shifted from the pump), and plot the resulting points in the complex plane. The point scatter then immediately illustrates the spread in amplitude and phase from shot to shot. Without the seed, we find that both the amplitude and phase variations are very large, filling a large region of the complex plane; whereas, when the correct seed is present, the scatter is compressed to a tiny fraction of the space.

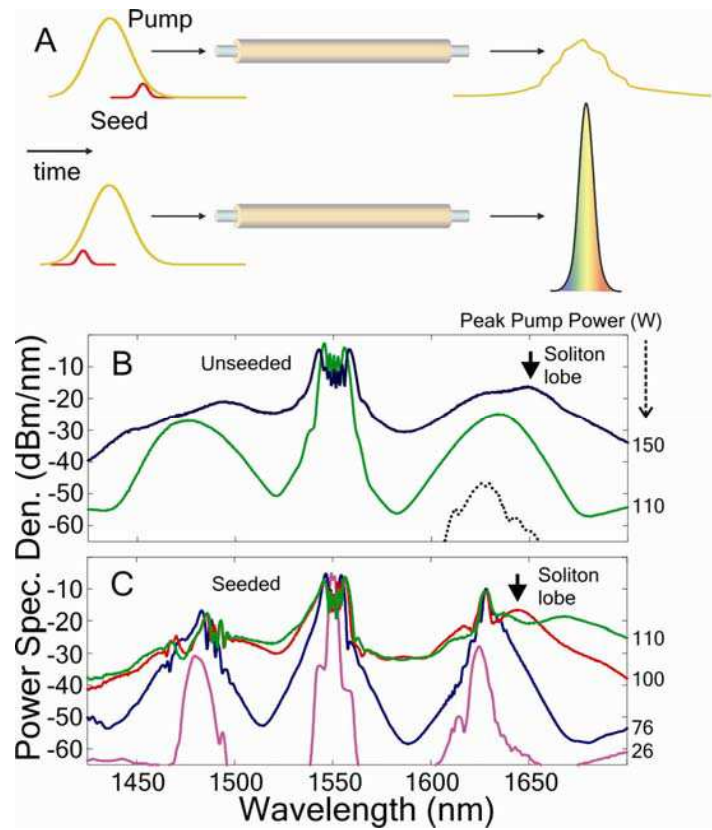

Figure 1 **Experimental measurement of supercontinuum (SC) spectra with and without the addition of a weak seed.** A) Illustration of the impact of a weak coherent seed on the input pulse. If the seed is properly timed (and has the correct center frequency), SC is readily produced; whereas, no SC is generated otherwise. B), C) Measured spectra without and with the properly timed seed, respectively, at the indicated pump (peak) power levels. The small arrows indicate the shedding of a soliton lobe from the red-shifted modulation-instability (MI) wing. When the seed is present, the soliton lobe is clearly discernable from the sharpened MI wing.

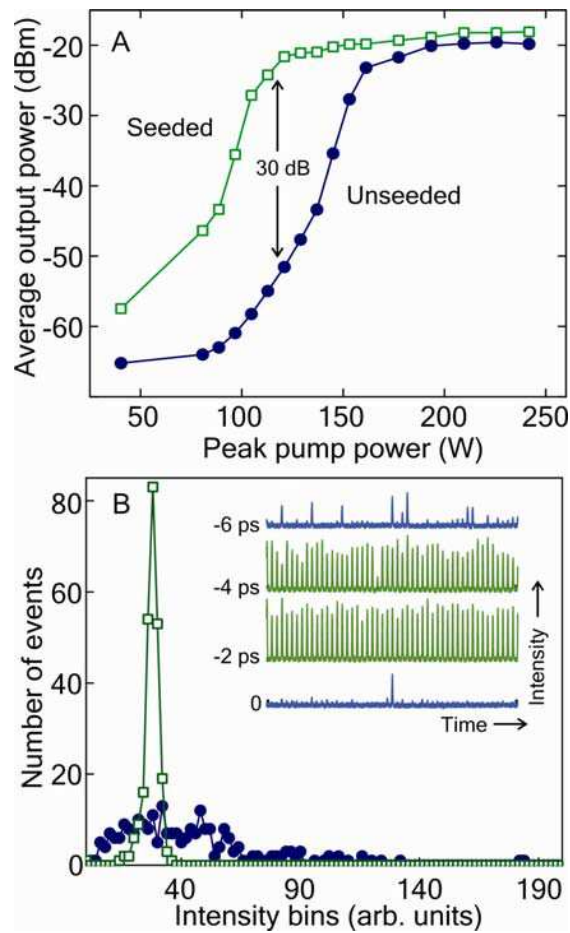

Figure 2 **Experimental measurement of SC output with and without stimulation.** A) Measured (filtered) SC power with and without seeding. SC generation exhibits a sharp threshold, which is dramatically reduced when the seed is added. B) Measured (filtered) SC pulse-to-pulse amplitude statistics with and without the seed (~100 events each). Pump power is adjusted (peak power ~130 W seeded, ~200 W unseeded) to keep the average SC output power constant. The histograms show that the weak seed dramatically reduces the statistical variation in the SC spectrum. The inset illustrates segments of the filtered SC pulse train for different time delays of the seed. Negative time delays correspond to an advanced seed pulse. When the seed is improperly timed, intense flashes of SC are rare; however, with proper timing, the intense events become the norm.

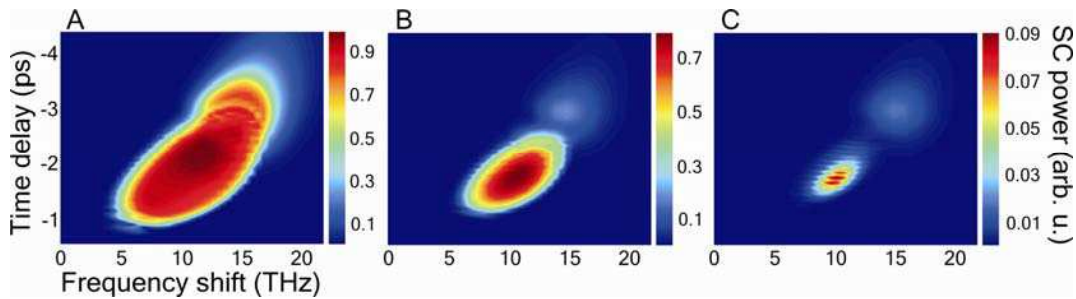

Figure 3 **Simulated maps of SC generation vs. timing and frequency shift of a weak seed pulse.** SC power integrated over red-shifted wavelengths ($\lambda$ >1680 nm) as a function of the time delay and frequency shift of the seed pulse for seed intensities: A) 0.01% of pump, B) 0.001% of pump, and C) 0.0001% of pump. The time delay is measured with respect to the center of the pump pulse; negative values correspond to the pump's leading edge. The input pump peak power is 130 W, the pump duration is 3.7 ps, and the seed pulse duration is 200 fs. The maps show a clear "sensitive spot" where the impact of the seed is maximized. As the seed power is reduced, the sensitive spot becomes smaller. Fringes are visible in the maps to varying degrees, and likely arise from spectral interference. Some additional structure is visible in the plotted region; however, a detailed analysis of these features is beyond the scope of the present study.

18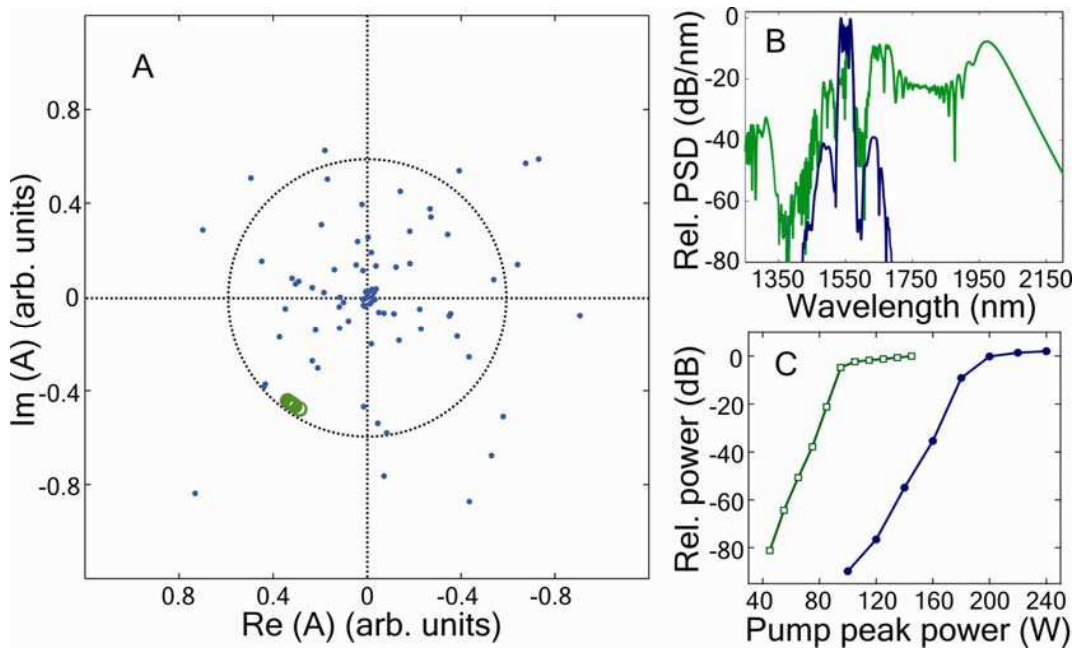

Figure 4 **Simulation of SC enhancement and stabilization with a weak seed.** A) Complex-plane scatter plot of the SC electric field amplitude $A$ ($\lambda$ = 1950 nm, $\Delta\lambda$ = 5 nm) from many independent events: 130 W pump peak power, seeded (green points, 40 events); 255 W, unseeded (blue points, 80 events). The 130 W pump generates negligible SC levels when unseeded. The 255 W pump generates appreciable SC, but the spectral amplitude and phase fluctuate tremendously from pulse to pulse. The seeded 130 W pump generates more SC than the unseeded 255 W pump, and the fluctuations are nearly eliminated at least ~400 nm from the pump (amplitude coefficient of variation: ~0.036 %; phase angle standard deviation: ~23 mrad). B) Single-shot seeded (green) and unseeded (blue) relative power spectral densities from a 130 W pump. C) Filtered (>1680 nm) SC power with (green squares) and without the seed (blue circles). In the latter case, we average many trials for each power level, as the unseeded output is unstable. The SC generation threshold is dramatically reduced when the seed is added.